# A Note on Fourier and the Greenhouse Effect

Joseph E. Postma
University of Calgary
October 7, 2015

Joseph Fourier's discovery of the greenhouse effect is discussed and is compared to the modern conception of the greenhouse effect. It is confirmed that what Fourier discovered is analogous to the modern concept of the greenhouse effect. However, the modern concept of the greenhouse effect is found to be based on a paradoxical analogy to Fourier's greenhouse work and so either Fourier's greenhouse work, the modern conception of the greenhouse effect, or the modern definition of heat is incorrect. The solution to this problem is not feigned to be given here.

## Fourier's Discovery

It is generally considered that Joseph Fourier discovered "the greenhouse effect". From the Wiki [1] article on Fourier:

> "In the 1820s Fourier calculated that an object the size of the Earth, and at its distance from the Sun, should be considerably colder than the planet actually is if warmed by only the effects of incoming solar radiation. He examined various possible sources of the additional observed heat in articles published in 1824 and 1827. While he ultimately suggested that interstellar radiation might be responsible for a large portion of the additional warmth, Fourier's consideration of the possibility that the Earth's atmosphere might act as an insulator of some kind is widely recognized as the first proposal of what is now known as the greenhouse effect.
>
> "In his articles, Fourier referred to an experiment by de Saussure, who lined a vase with blackened cork. Into the cork, he inserted several panes of transparent glass, separated by intervals of air. Midday sunlight was allowed to enter at the top of the vase through the glass panes. The temperature became more elevated in the more interior compartments of this device. Fourier concluded that gases in the atmosphere could form a stable barrier like the glass panes. This conclusion may have contributed to the later use of the metaphor of the 'greenhouse effect' to refer to the processes that determine atmospheric temperatures. Fourier noted that the actual mechanisms that determine the temperatures of the atmosphere included convection, which was not present in de Saussure's experimental device."

Of course, the postulate that "gases in the atmosphere could form a stable barrier like the glass panes" is nonsensical, and in as much does not support the concept or a discovery of an atmospheric greenhouse effect. As pointed out however, modern usage of the phrase 'greenhouse effect' has changed and is in fact "metaphorical" in reference to the phenomenon Fourier was discussing at the time.

## de Saussure's Vase

Horace-Bénédict de Saussure [2] created a simple device to demonstrate the type of greenhouse effect we find in a botanist's glass greenhouse, which had results that significantly impressed upon Fourier [3]:

> "We owe to the celebrated voyager M. de Saussure an experiment which appears very important in illuminating this question. It consists of exposing to the rays of the sun a vase covered by one or more layers of well transparent glass, spaced at a certain distance. The interior of the vase is lined with a thick envelope of blackened cork, to receive and conserve heat. The heated air is sealed in all parts, either in the box or in each interval between plates. Thermometers placed in the vase and the intervals mark the degree of heat acquired in each place. This instrument has been exposed to the sun near midday, and one saw, in diverse experiments, the thermometer of the vase reach 70, 80, 100, 110 degrees and beyond (octogesimal division). Thermometers placed in the intervals acquired a lesser degree of heat, and which decreased from the depth of the box towards the outside.
>
> "The effect of solar heat on the air trapped by the transparent envelopes has been observed long since. The apparatus which we have just described has the objective of taking the heat acquired to its maximum, and above all to compare the solar effect on a high mountain to that taking place on the plain beneath. This observation is principally remarkable for the sound and extensive results that the inventor has been able to make: it has been repeated several times at Paris and Edinburgh, and has given analogous results.
>
> "The theory of this instrument is easy to understand. It suffices to remark, firstly that the heat acquired is concentrated, because it is not immediately dissipated by the renewing of the air; secondly that the heat emanating from the sun has different properties to that of IR. The SW is nearly completely transmitted by the panes of glass [in all capacities] to the bottom of the box. It heats the air and the walls which contain it: then their heat thus communicated ceases to be luminous; it only conserves the properties of IR. In this state it cannot freely traverse the panes of glass which cover the vase; it accumulates more and more in a space enveloped in poorly conducting material, and the temperature rises until the incoming heat is exactly compensated for by that which dissipates. One could verify this explanation, and render the consequences more sensible, if one varied the conditions, in employing glasses coloured or blackened, and if the spaces which hold the thermometers are emptied of air. When one examines this effect via calculus, one finds results entirely conforming with those given by observations. It is necessary to consider attentively this order of facts and the results of calculus when one wishes to know the influence of the atmosphere and the waters on the temperature of the earth.

"In effect, if all the levels of the air of which the atmosphere is formed were to retain their density and transparency, and lose only their mobility, this mass of air thus becoming solid, being exposed to the rays of the sun, would produce an effect of the same type as that which one has just described. The heat, arriving as SW at the surface of the earth, would suddenly lose entirely the faculty which it had of traversing diaphanous solids; it would accumulate in the lower levels of the atmosphere, which would thus acquire elevated temperatures. One would observe at the same time a diminution of the degree of heat acquired, above the surface of the earth. The mobility of the air which moves rapidly in all directions and which rises when heated, the radiation of IR in the air, would diminish the intensity of the effects which would take place under an transparent and solid atmosphere, but would not entirely remove these effects. The decrease of heat in the higher regions of the air does not cease to take place; it is thus that the temperature is augmented by the interposition of the atmosphere, because the heat finds fewer obstacles in penetrating the air, when it is SW, than in repassing when converted into IR."

The idea that levels of atmosphere could "lose only their mobility, this mass of air thus becoming solid" has already been remarked upon. That is, any conception or remarks from Fourier postulating that horizontal slices of the atmosphere can act as solid barriers is nonsensical and such statements of course do not support a discovery of or a concept of an atmospheric greenhouse effect.

Of the maximum temperature observed inside de Saussure's device, it is said to be 110 degrees Celsius "and beyond". We are not told how much beyond, but we are indicated as to 70, 80, and 100 degrees as other representative values, and so we will later outline the methodology required to properly recreate such a device and perform the requisite experiment, to see how warm it can actually get. For now, we can employ the 1$^{st}$ Law of Thermodynamics to estimate the maximum temperature of de Saussure's device and any other phenomena following it.

Temperature of a "Greenhouse Box"

The rate of change of temperature of a surface is proportional to the rate of heat flow through the surface, where the heat flow is given by the difference between the rate of energy input from a power source (such as the sun & its sunlight) absorbed on the surface, and the rate of energy output from the surface factored for relevant physical characteristics such as emissivity. In this context, the energy parameters are fluxes (F):

$$\frac{dT}{dt} \propto F_{in} - F_{out} \qquad (1)$$

Our scenario is for a thin plane surface with radiative energy transfer only, using a greenhouse box which is insulated to convective and conductive energy loss; the only loss of energy will occur on the "radiatively open" exterior which faces the power source. The radiant flux $F_{out}$ is given by the surface's temperature in the Stefan-Boltzmann equation factored for its emissivity, while $F_{in}$ is given by the radiant solar flux at the relevant location on the Earth factored for atmospheric extinction and the absorptivity of the surface. (See Appendices A, B, and C for convenient Matlab function scripts for calculating the top-of-atmosphere solar flux from the sun for any given input date and location.)

The definition of thermal equilibrium is such that $dT/dt = 0$, and thus at thermal equilibrium $F_{in} = F_{out}$ which allows for the equilibrium temperature to be directly solved:

$$(1-\varepsilon)(1-\alpha) F_{TOA} = \mu \sigma T_{surf}^4 \qquad (2)$$

In the "diverse experiments" Fourier references in regards to de Saussure's device, we would expect that the results with the highest temperature arose from those conditions which were most conducive to the collection of solar energy. Thus, for an atmospheric total extinction of solar light of $\varepsilon \sim 14\%$ when the sun is near solar noon on a clear day and at some altitude up a mountain, a maximum TOA flux of $F_{TOA} = 1415$ W/m², a surface emissivity of $\mu = 96\%$ and surface albedo of $\alpha = 96\%$ (such as for charcoal, given that de Saussure's device utilized "blackened cork" for the absorptive medium), then the equilibrium temperature is 109.8 degrees Celsius, which therefore provides the theory explanation for de Saussure's observation of the device reporting a maximum of about 110 degrees Celsius. The experimental uncertainty is unknown for de Saussure's observations but clearly the calculated prediction and his empirical result are satisfyingly similar. Variations of the atmospheric extinction, the solar altitude, the physical altitude of the device in atmosphere, time of year, etc., explains the lower temperature results, or higher as the case may be (for example, if the extinction is reduced by venturing higher atop a mountain, or if emissivity or albedo is reduced, etc.). We will see later how to reconstruct such a thermal response device and how to take account of all such experimental factors.

Fourier remarks that in this device "the heat acquired is concentrated". By today's standards this is relatively inaccurate terminology, as we read on "heat" in Thermodynamics [4]:

> "Heat is defined as the form of energy that is transferred across a boundary by virtue of a temperature difference or temperature gradient. Implied in this definition is the very important fact that a body never contains heat, but that heat is identified as heat only as it crosses the boundary. Thus, heat is a transient phenomenon. If we consider the hot block of copper as a system and the cold water in the beaker as another system, we recognize that originally neither system contains any heat (they do contain energy, of course.) When the copper is placed in the water and the two are in thermal communication, heat is transferred from the copper to the water, until equilibrium of temperature is established. At that point we no longer have heat transfer, since there is no temperature difference. Neither of the systems contains any heat at the conclusion of the process. It also follows that heat is identified at the boundaries of the system, for heat is defined as energy being transferred across the system boundary."

Thus, what Fourier is actually referring to in the thing which is "acquired and concentrated" is *energy*, with the heat input actually going to zero when thermal equilibrium is achieved at maximum temperature. The source of energy being concentrated, though not stated quite directly, is of course that provided by a power source, which in this case is the Sunlight input to the device. Firstly, the stacked glass panes prevent the convection of cooler air coming in and replacing the heated air inside the interior chambers. The interior maximum air temperature would be reduced from its full potential if it were continually replaced via convection by cool air from the environment, and so by holding a defined parcel of air in place it will thus attain the maximal temperature of the surface it is in contact with where solar radiation is being converted into a material temperature due to its absorption on the surface.  Additionally, if the glass panes are opaque to long-wave thermal radiation, then they will be heated by that radiation outgoing from the interior surface according to their absorptivity and emissivity and this would serve to increase the speed at which the maximum interior air temperature is achieved.  The final, outward exterior of such stacked opaque glass panes then becomes the surface from which radiant thermal energy is emitted, and the emissivity of the glass will of course have an effect on the final temperature they themselves achieve.

## The Modern, or "IPCC", Greenhouse Effect

We will label the modern conception of the atmospheric greenhouse effect as the "IPCC greenhouse effect", where IPCC of course stands for the Intergovernmental Panel on Climate Change, and its greenhouse effect as the one it subscribes to. Generally, it is the greenhouse effect which is used in climate science and which underlies the climate models and climate assessments and which is dependent upon greenhouse gases. Recall, as per the Wiki quote on Fourier, that the modern greenhouse effect is "metaphorical" in relation to the one discussed by Fourier, and so the difference therein is what we shall discuss now.

For example, from the Department of Atmospheric Sciences at the University of Washington [5] (see Appendix D for a sample listing of identical presentations from other institutions, which establishes that this indeed the accepted modern greenhouse mechanics of energy transfer in the atmosphere or a device which replicates such mechanics):

> "…the top layer of the atmosphere must emit 239.7 W/m² of infrared radiation to space (same amount of solar radiation that enters the atmosphere: what goes in must go out). The bottom layer of the atmosphere will emit an equal amount downward to the surface of the planet. Hence, for thermal equilibrium, the surface of the planet must emit enough radiation to balance not only the amount it receives from the sun (239.7 W/m²), but also what it receives in the form of downward infrared radiation from the atmosphere 239.7 W/m²). Hence, its emission must match 239.7+239.7 = 479.4 W/m². Applying the Stefan-Boltzmann law: constant x T⁴ = 479.4 W/m². We thus calculate T = 303 K."

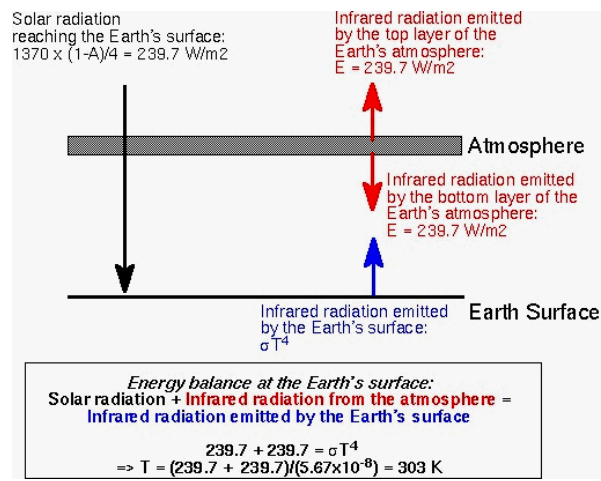

*Figure 1: From http://www.atmos.washington.edu/2002Q4/211/notes_greenhouse.html*

In this modern definition of the atmospheric greenhouse effect we have something different occurring than in Fourier's greenhouse effect. Instead of the solar heat forcing being maximized to its own full potential, this metaphorical or analogous version of the greenhouse effect has radiation from the atmosphere adding to the solar forcing such as to increase the total radiant flux driving the generation of temperature on the absorbing surface.

The analogy to Fourier's greenhouse effect, i.e. the metaphor, is that the glass panes of de Saussure's device could be imagined to do something similar. For example, if the glass pane were opaque to the longwave radiant thermal energy emitted from the interior back-surface, which was absorbing $(1-\varepsilon)(1-\alpha)F_{TOA} = 1168.2 \text{ W/m}^2$ shortwave radiant energy from the sun from the previous example, then an equal addition of this flux would impinge the back-surface due to longwave emission from the pane resulting in a final temperature of 182.2 degrees Celsius for the back-surface. The longwave radiant thermal emission from the glass pane back to the back-surface is called "backradiation" in modern greenhouse effect science, and it represents the mechanism by which temperatures are thought to be amplified in a greenhouse and in the atmosphere within that context.

In fact, whether or not the glass pane of this device is opaque to longwave radiation, a greenhouse effect of the modern type could be observed because either: A) the glass is opaque to longwave radiation and thus produces its own modern greenhouse effect on the interior of the box, or, B) the glass is transparent to longwave radiation and thus receives the longwave emission from the atmosphere which produces the atmosphere's own modern greenhouse effect. And if the pane is semi-transparent to longwave, then it will have a combination of its own internal modern greenhouse effect and a contribution of the atmosphere's modern greenhouse effect. If we apply the additional long-wave flux of 324 W/m² absorbed by the surface from the atmosphere [6] to the shortwave solar input forcing flux into de Saussure's device, then a final equilibrium temperature of 133.9 degrees Celsius could be found.

The Greenhouse Box

      We assume that the laws of physics are universal, or are at least locally ubiquitous to some certain extent, and thus whether radiation is emitted from an atmosphere or a glass pane, that if the "backradiation" emitted from an atmosphere causes a modern greenhouse effect on a heated surface beneath it, then the same physics should hold for a glass pane emitting backradiation to a heated surface underneath it such as in a "de Saussure vase" greenhouse device, the relevant physics being that the secondary longwave radiant thermal emission adds to the primary incoming shortwave radiant forcing of the power source.  And in fact, whether or not the glass panes of a "de-Saussure vase" are opaque or transparent to longwave radiation, a modern greenhouse effect will be measureable in such a device at the surface of the Earth since it either creates its own modern greenhouse effect internally, or, it gets acted upon by the atmosphere's modern radiative greenhouse effect.

      The existing empirical results of Fourier and de Saussure are not supportive of the modern atmospheric IPCC greenhouse effect.  As we have seen, Fourier and de Saussure reported a maximum temperature of $110^0$C in their device which is simply equal to the value expected given the primary input forcing from Sunlight, whereas the IPCC greenhouse effect indicated a significantly higher temperature to be found of at least $134^0$C and as high as $182^0$C with only a single additional pane of longwave-opaque glass.  Such additional forcing to higher temperature should have easily been observed and reported by Fourier and de Saussure, particularly with the multiple panes of glass in their device.

      If heat is defined as the "energy transfer across a boundary by virtue of a temperature gradient", and the direction of this transfer is "down the gradient", i.e. from higher to lower temperature, then it is unclear how thermal energy held within a cooler atmosphere or a cooler glass pane(s) could transfer by heat to a warmer primary surface that is itself the original source of heat.  Given that this is a general definition, its resultant restrictions must apply to all the modes of heat transfer being conductive, convective, and radiative (or just physical and radiative).  "Secondary heating", i.e., where heat is transferred from cool back to warm after warm has heated the cool, seems to be an incommensurate postulate to the definition of heat.  It also seems like such a mechanism would be a positive-feedback runaway process.  The modern IPCC radiative greenhouse effect is not supported by Fourier and de Saussure's existing empirical results, nor the by the modern definition of heat.

A different explanation could be employed in that the heat from the glass panes or atmosphere provides resistance to heat flow out of the surface underneath, rather than there being any actual transfer of heat from cool to hot, and thus the primary surface absorbing the primary energy is amplified in temperature in order to overcome the resistance provided by the backradiation from the cooler layers. However, such a concept of "heat flow resistance caused by the presence of a cooler temperature thus leading to a higher source temperature" is not readily discoverable in thermodynamic textbook sources, and, it would in any case produce the same result that a higher temperature should have been empirically measured by Fourier and de Saussure. Photons are bosons and it is not as if they push against each other until enough "temperature force" is acquired to squeeze them back out of the device.

In any case, many people believe in the IPCC modern radiative greenhouse effect and there is an undeniable scientific consensus as to its veracity. Consequently, it must be correct. Therefore, the simplest thing to do is to perform an appropriate empirical experiment on the IPCC radiative greenhouse mechanism, and de Saussure's "glass vase" device is an excellent apparatus for such a task since, as we have assumed, the laws of physics are universal. It is simply a box with a known low albedo interior in both shortwave and longwave, heavily insulated on all sides except for the side which will point toward the sun, this side having either long-wave transparent or long-wave opaque pane(s) of glass.

The quantities to be known and measured are for example:
- the actual solar flux being absorbed by the back-surface inside the box
- the albedo and emissivity of the absorbing back-surface
- the temperature of the back-surface and the panes of glass
- the emissivity of the glass panes will have an effect on their own temperature and the long-term temperature of the entire box

The idea is of course simply to test whether the temperature of the absorbing back-surface inside the box is determined by the primary solar heat forcing, or, by the primary solar contribution plus a secondary back-heat forcing provided by the passive elements of the glass panes and/or atmosphere. The temperature will need to be measured vs. time, and as the back-surface comes to equilibrium with its heat input then the plot will asymptote to its equilibrium value.

The Scenario

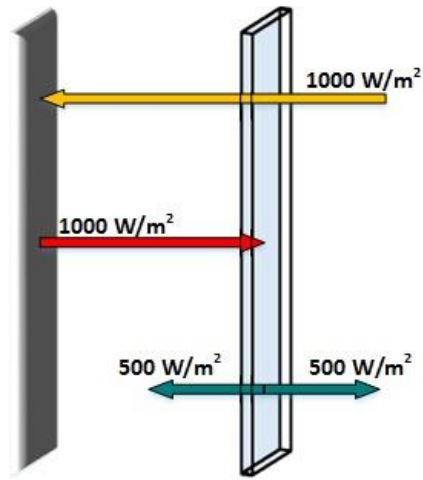

*Figure 2: How do we solve this problem?*

The modern IPCC radiative greenhouse effect begins with the problem depicted in the image above: Shortwave radiation passes through a substance (in this case, a pane of glass or an atmosphere) of which it is transparent, and is then absorbed by another surface further on. The shortwave-absorbing surface is insulated on its backside. The shortwave surface then thermally radiates back out in one direction only, but at longwave frequencies to which the other substance or layer is now opaque. Thus, the other layer absorbs the longwave radiation from the shortwave-absorbing surface. At this point, the longwave layer emits its own thermal radiation, but it must emit it in two directions because it is not insulated on either side. The longwave layer thus emits back towards the shortwave surface and also back to outer space, and so to conserve energy it only emits half of the energy it absorbs to either of those directions. At this point, total energy is not conserved to the outside of the system because only half of the input energy is making it back out to outer space. So, how do we proceed?

If we consider that this system does not actually start off as depicted in Figure 2, but would be a differential calculus system beginning initially with zero outputs, and if we maintain the split-by-two property of the secondary layer, then the end-result would be 1000 W/m² emitted outward by the secondary layer with 2000 W/m² being emitted by the back-surface, as in the next figure.

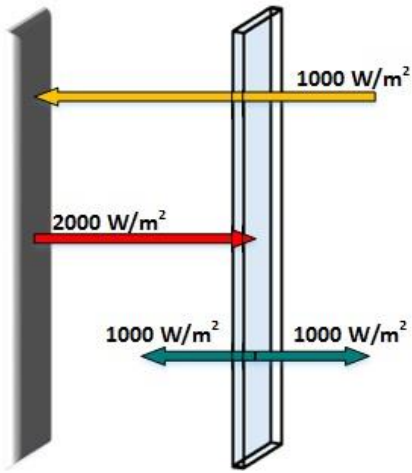

*Figure 3: What appears to be a reasonable solution.*

In a de Saussure IPCC greenhouse device with multiple layers, then the end result would take on the following energy densities:

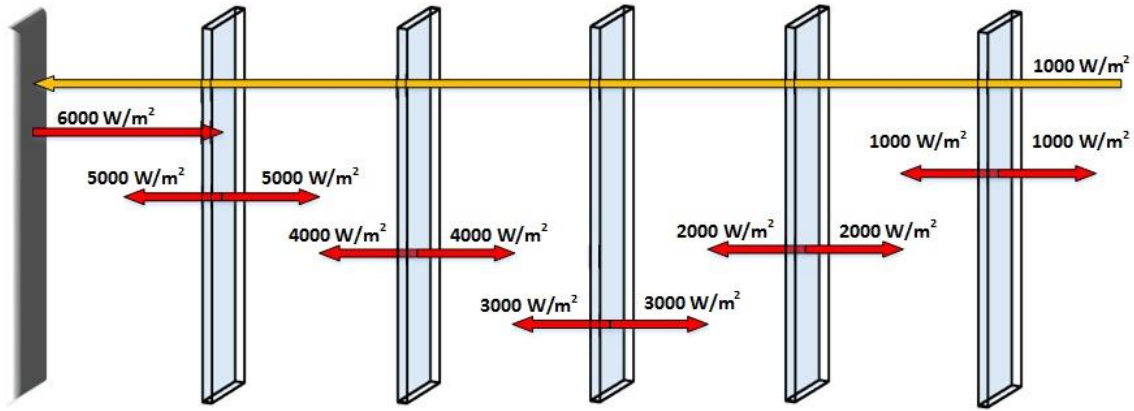

*Figure 4: What appears to be a very practically useful solution.*

A 5-layer de Saussure IPCC greenhouse device would result in a back-surface energy flux of 6,000 W/m², which is 570⁰K or 297⁰C or 566⁰F via the Stefan-Boltzmann Law. This seems to be a very practically useful result as it indicates that a primary radiant heating source (1000 W/m² of solar energy in this case) can be concentrated or amplified to temperatures far warmer than the equivalent temperature of the primary initial radiant heat source itself. Indeed, in theory it would work better than even a magnifying glass or focusing mirror and would not be limited by the effective temperature of the source spectrum since there is no limitation in these mechanics on the thermal properties of the primary heat source. The device de Saussure used was said to have multiple panes of glass, and so the effect predicted by the modern IPCC greenhouse effect should have been readily apparent.

If the textbook definition of heat flow is correct, and if it has any relevance to this scenario, that is, if heat only flows down the temperature gradient and heat does not resist its own spread when it flows into and heats something, and, positive heat input is what is required to raise a temperature, then one might be tempted to conclude that Figure 5 below may represent the equilibrium condition with the primary input. Since heat never flows from the layer to the back-surface because the layer is always cooler than the back-surface, or at most the same temperature, then the layer doesn't lose any energy to the inside of the compartment and only loses it to the outside. Stacked panes of glass would then be considered to function as insulation, rather than as heat amplifiers. Such a function as insulation is not the modern IPCC radiative greenhouse effect of climate science.

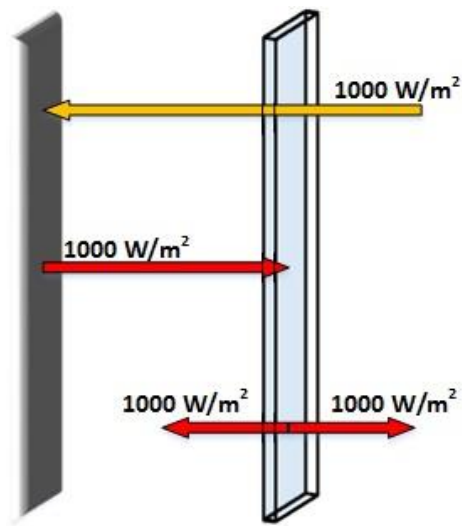

*Figure 5: This solution is only true if modern textbooks defining heat flow have "got it right".*

The radiation flux vector terms in the above figure for the back-surface (left) and the pane (right) can simply be replaced by the scalar temperatures of the surfaces (91$^0$C), and when the surfaces are viewed with their temperatures then it may be easier to conclude that no heat flows between them. Energy may be shared between the surfaces, but energy in the form of heat, according to the definition of heat, does not transfer. Without transfer of heat, then no temperatures can be affected due to any supposed secondary sharing of energy between the back-surface and glass pane. In this case, only the primary input represents a forcing for temperature, and secondary radiation or "backradiation" cannot serve as a temperature forcing amplifier.

## Conclusion

Secondary radiation, or "backradiation", acts as a temperature amplifier in the modern IPCC conception of the greenhouse effect of climate science, and so the effect should be measurable in any scenario where backradiation has to exist, such as is a de Saussure "glass vase device", or any botanist's greenhouse.  That is, the temperature inside these devices should exceed that of the temperature of any primary heat forcing, such as sunlight for example.  The backradiation effect of the modern IPCC radiative greenhouse effect allows, in theory, any temperature to be achieved that one wishes by simply adding the requisite number of panes of glass, sufficiently engineered, to the greenhouse device.  These mechanics should allow for the temperature of the primary heat forcing power source input itself to be increased at will.  For example, if one had a coal-powered steam engine, then simply wrapping the coal combustion chamber with extra layers of steel (with gaps in between them) would increase the temperature of the combustion since steel is opaque to longwave radiation - this would send backradiation which would then increase the combustion temperature beyond that provided by the molecular transmutation of the coal.  The increases in efficiency in the utilization of the energy chemically released from a lump of coal would have been quite useful.

> "Once you have assumed the wrong ontology and epistemology, everything you subsequently say is automatically in error."
>
> -Hockney, Mike (2015-06-02). The War of the Ghosts and Machines (The God Series Book 28) (Kindle Locations 655-656). Hyperreality Books. Kindle Edition.

# Appendix A: Matlab Function for Calculating Solar Altitude, Airmass, and Distance

```
function [alt varargout] = Solar_Altitude(JD,latitude,longitude)
%[Altitude(deg), Airmass(opt.1), Distance(opt.2, meters)] =
%Solar_Altitude(Julian Day, Latitude, Longitude);
%Returns solar altitude and optionally the airmass & distance, accurate to
%~0.01 degrees between 1950 and 2050.
%Julian Day can be a vector, but the other parameters should be scalars.
%If solar altitude is below horizon, airmass will report airmass for 90
%degrees, equal to 31.7349.

%Number of days since JD 2451545.0 = 2000 UT 12:00:00
n = JD - 2451545.0;%days

%Mean longitude (corrected for aberration);
L = rem(280.460 + 0.9856474*n,360);%deg

%Mean anomaly;
g = rem(357.528 + 0.9856003*n,360);%deg

%Ecliptic Longitude
lambda = L + 1.915*sind(g) + 0.020*sind(2*g);%deg

%Obliquity of Ecliptic
eps = 23.439 - 0.0000004*n;%deg

%Solar Right Ascension
f = 180/pi;
t = tand(eps/2).^2;
alpha = rem(lambda - f*t.*sind(2*lambda) + ...
(f/2)*(t.^2).*sind(4*lambda),360);%deg

%Solar Declination
delta = asind(sind(eps).*sind(lambda));%deg

%West longitude of observatory in hours
WLO = longitude/15;%hrs

% Greenwhich Mean Sidereal Time at JD
GMST = rem(18.697374558 + 24.06570982441908*(JD - 2451545.0),24);%hrs

% Local Sidereal Time at JD and longitude
LST = GMST - WLO;%hrs

%Solar local hour angle
ha = LST - alpha*12/180;%hrs
ha = ha * 180/12;%degrees

%Solar Altitude, degrees
alt = asind(sind(latitude).*sind(delta) + ...
      cosd(latitude).*cosd(delta).*cosd(ha));
```

```matlab
alt = alt(:);

if nargout >= 2    %Solar Airmass
    %Zenith angle
    zt = (90 - alt); zt(zt > 90) = 90;
    %Airmass of target: Young, A. T. 1994. Air mass and refraction. Applied
    %Optics. 33:1108-1110.
     X = ( 1.002432*cosd(zt).^2 + 0.148386*cosd(zt) + 0.0096467 ) ./ ...
(cosd(zt).^3 + 0.149864*cosd(zt).^2 + 0.0102963*cosd(zt) + 0.000303978);
     varargout(:,1) = X{:};
end

if nargout >= 3    %Solar Distance
    d = (1.00014 - 0.01671*cosd(g) - 0.00014*cosd(2*g))*149597870691; %149597870691 = 1au (meters)
    varargout(:,2) = d{:};
end
```

# Appendix B: Matlab Function for Calculating Top of Atmosphere Solar Flux

```matlab
function res = TOA_Flux(JD, lat, long)
%JD = Julian Date
%lat = Latitude of location, degrees
%wlong = West Longitude of location, degrees

%constants
sbc = 5.67e-8;   %Stefan-Boltzmann Constant
SeffT = 5778;    %Solar Effective Temperature; K, value from Gray
RSun = 6.96e8;   %Radius of Sun; meters

%Solar altitude, airmass, and distance
[Sol_Alt Sol_X Sol_Dist] = Solar_Altitude(JD,lat,wlong);

%Flux at TOA
res = sbc*(SeffT^4)*(RSun./Sol_Dist).^2;
```

## Appendix C: Matlab Function for Calculating the Julian Date

```matlab
function JD = date2JD(date,utime)
%jd = date2JD(date,utime);
%Convert Calendar Date & Universal Time to Julian Day Number
%-date is the year-month-date string = '2000-03-20' for example
%-utime is the universal time string = '04:15:16' for example
%-utime can also be a decimal-valued hour time = 4.25444444 for example
%-date & utime can be a scalar string value for single-value returns
%-date & utime can be equal-length cell arrays of string/numeric values for
%vector returns

%#days in each month for a LEAP year
monthsly  = [0 31 29 31 30 31 30 31 31 30 31 30];
%#days in each month for a non-LEAP year
monthsnly = [0 31 28 31 30 31 30 31 31 30 31 30];
%reference Julian Date on 2000-01-01 @ 0.0hrs UT
JD2000UT00 = 2451544.5;

if ~iscell(date)%then date is a scalar single string value
    date = {date};
    utime = {utime};
end

JD = zeros(length(date),1);

for i = 1:length(date)
    indsd = strfind(date{i},'-');%indices of dash in date
    indst = strfind(utime{i},':');%indices of colon in time
    
    %check date format
    if (length(indsd) ~= 2 || indsd(1)~=5 || indsd(2)~=8 || ...
            length(date{i}) ~= 10)
        error(['Value for DATE argument not correct format at entry ',...
            num2str(i),'; Value = ', date{i}]);
        return;
    end
    
    %check time format
    if ~isnumeric(utime{i})
        if (length(indst) ~= 2 || indst(1)~=3 || indst(2)~=6  ...
                || length(utime{i}) ~= 8)
            error(['Value for TIME argument not correct format at entry ',...
                num2str(i),'; Value = ', utime{i}]);
            return;
        end
    end
    
    %get year, month, day
    year = str2double(date{i}(1:4));
    month = str2double(date{i}(6:7));
    day = str2double(date{i}(9:10));
    %number of leap years since & including 2000
    Nly = floor((year-1-2000)/4) + 1;
```

```matlab
    %number of non-leap years since 2000
    Nnly = year - 2000 - Nly;
    
    %is this year-date a leap year?
    if rem(year-2000,4) == 0
        months = monthsly;
    else
        months = monthsnly;
    end
    
    Ndays = 366*Nly + 365*Nnly + sum(months(1:month)) + day - 1;
    
    t = 0;
    
    if ~isnumeric(utime{i})
        t = str2double(utime{i}(1:2))/24 + ...
                str2double(utime{i}(4:5))/24/60 + ...
                    str2double(utime{i}(7:8))/24/60/60;
    else
        t = utime{i}/24;
    end
    
    JD(i) = JD2000UT00 + Ndays + t;
end
```

## Appendix D: Consensus Listing of the Modern Atmospheric Greenhouse Mechanism

Notable quotes have been underlined by the author:

**Harvard University**

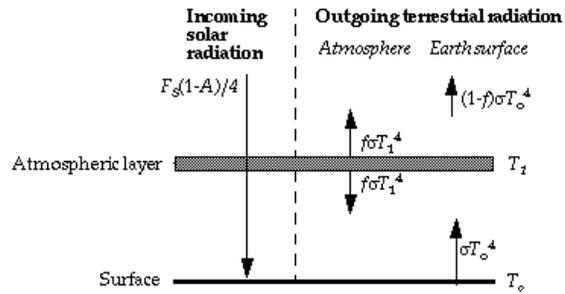

http://acmg.seas.harvard.edu/people/faculty/djj/book/bookchap7.html

**Pennsylvania State University**

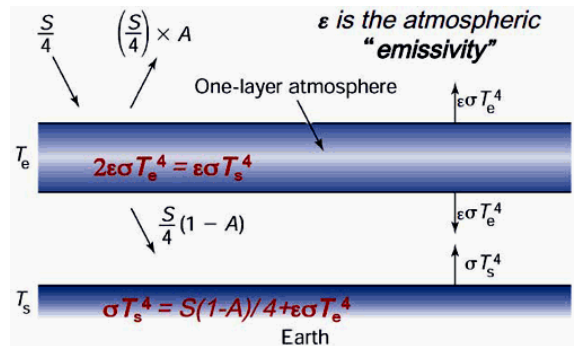

https://www.e-education.psu.edu/meteo469/node/198

**University of Chicago**

Found in Chapter 3, lecture 5 video lecture: *The Greenhouse Effect*.

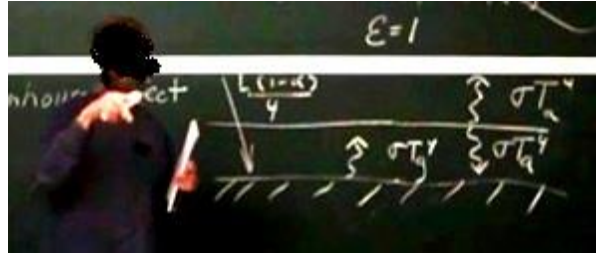

http://mindonline.uchicago.edu/media/psd/geophys/PHSC_13400_fall2009/lecture5.mp4

**University of Washington's Department of Atmospheric Sciences.**

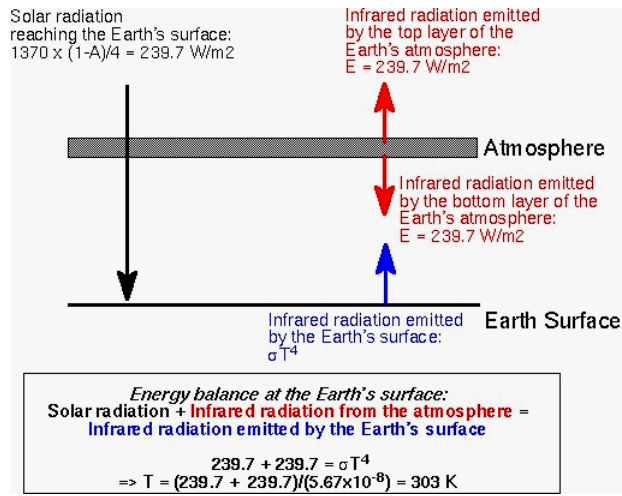

http://www.atmos.washington.edu/2002Q4/211/notes_greenhouse.html

*Columbia University*

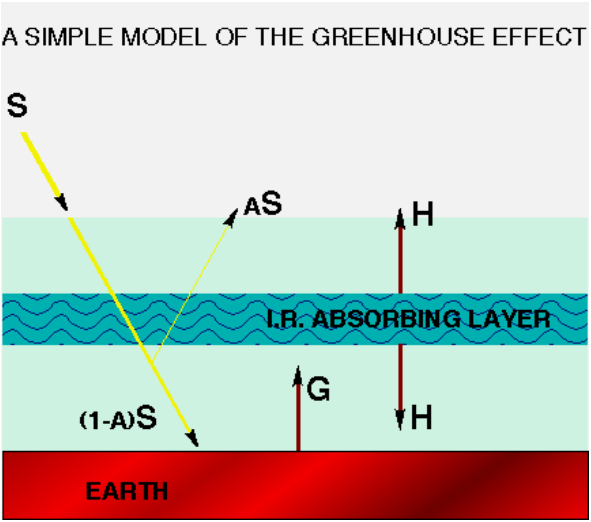

http://eesc.columbia.edu/courses/ees/climate/lectures/gh_kushnir.html

**Kiehl & Trenberth, "Earth's Annual Global Mean Energy Budget," Bulletin of the American Meteorological Society**

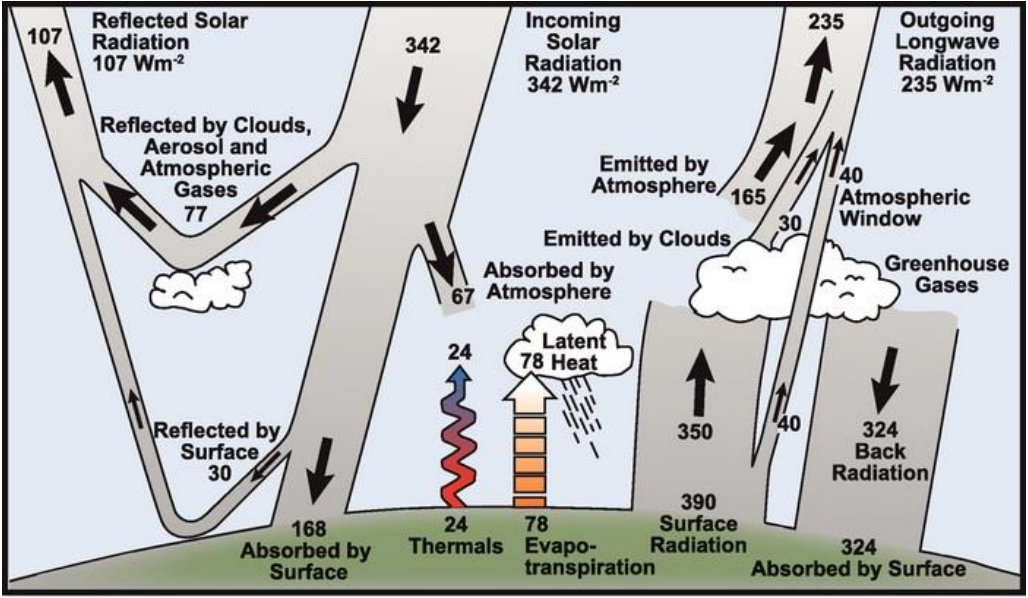

### NASA

"Why is this process called "The Greenhouse Effect?" The Sun heats the ground and greenery inside the greenhouse, but <u>the glass absorbs the re-radiated infra-red and returns some of it to the inside</u>."

http://www-istp.gsfc.nasa.gov/stargaze/Lsun1lit.htm

### Hunan University, China

"Light from the sun includes the entire visible region and smaller portions of the adjacent UV and infrared regions. Sunlight penetrates the atmosphere and warms the earth's surface. Longer wavelength infrared radiation is radiated from the earth's surface. A considerable amount of <u>the outgoing IR radiation is absorbed by gases in the atmosphere and reradiated back to earth</u>. The gases in the atmosphere that act like glass in a greenhouse are called greenhouse gases."

http://jpkc.lzjtu.edu.cn/hjhx/jpkc/7.ppt

### Appalachian State University, North Carolina

"Our atmosphere is a selective filter since it is transparent to some wavelengths and absorbs others. The greenhouse effect occurs when the energy absorbed is not all radiated because of the filtering of the atmosphere. <u>Some of the earth's radiated energy is reflected back to the surface</u>. Consequently the earth's atmosphere has an increased temperature. <u>This process is much like the action of glass in a greenhouse</u>."

http://www.physics.appstate.edu/courses/FirstExamReview.rtf

### The University of the Western Cape, South Africa

"A greenhouse is made entirely of glass. When sunlight (shortwave radiation) strikes the glass, most of it passes through and warms up the plants, soil and air inside the greenhouse. As these objects warm up they give off heat, but these heat waves have a much longer wavelength than the incoming rays from the sun. <u>This longwave radiation cannot easily pass through glass, it is re-radiated into the greenhouse, causing everything in it to heat up</u>. Carbon dioxide is the pollutant most responsible for increased global warming."

http://www.botany.uwc.ac.za/envfacts/facts/gwarming.htm

**The Institute for Educational Technology, Italy**

"*Just as it happens in a greenhouse where the function carbon dioxide performs in the atmosphere is played by glass-rafters, the sun's energy arrives down at the earth, where it is partially absorbed and partially reflected. Such reflected heat, however, is reflected again, by glass as for the greenhouse, by carbon dioxide as for the atmosphere, down on earth: it is as if a part of the heat were entrapped, thus determining a growth of temperature on the ground.*"

http://www.itd.cnr.it/ge8/rivista/inglese/num_2/galil3.htm

**The Austrian JI/CDM- Programme**

"*The Earth's atmosphere is comparable to a glass roof of a greenhouse: the short-wave solar radiation passes through nearly unhindered and warms the Earth's surface. From the Earth's surface, the short-wave radiation is partly absorbed and partly reflected back as long-wave thermal radiation. However, this long-wave thermal radiation cannot pass the atmosphere unhindered due to the greenhouse gases but is partly reflected back again to the Earth's surface.*"

http://www.ji-cdm-austria.at/en/portal/kyotoandclimatechange/ourclimate/greenhouseeffect/

**U.S. Department of the Interior, U.S. Geological Survey**

"*The gases that encircle the Earth allow some of this heat to escape into space, but absorb some and reflect another portion back to the Earth. The process is similar in Mountain View, only, the greenhouse there is made of glass instead of gas.*"

http://hvo.wr.usgs.gov/volcanowatch/1998/98_10_22.html

**RealClimate**

"*The factor of two for the radiation emitted from the atmosphere comes in because the atmosphere radiates both up and down.*"

http://www.realclimate.org/index.php/archives/2007/04/learning-from-a-simple-model/

**ThinkQuest Education Foundation**

"*In a greenhouse, heat from the sun enters the glass. The heat in the form of infra-red light bounces and heads back up towards the glass. The glass then allows only some of this heat to escape, but reflects back another portion. This heat remains bouncing within the greenhouse. In the case of planet Earth, there is no glass, but there is an atmosphere which retains heat or releases heat.*"

http://library.thinkquest.org/11353/greenhouse.htm

**UK government website:**

"*After gas molecules absorb radiation, they re-emit it in all directions. Some of the infrared radiation absorbed by gases in the atmosphere is therefore re-radiated out towards space and eventually leaves the atmosphere, but some is re-radiated back towards the Earth, warming the surface and lower atmosphere (illustrated by the 'back radiation' term in Figure 2). This warming is known as the greenhouse effect and the gases that are responsible for it are known as greenhouse gases.*"

http://www.bis.gov.uk/go-science/climatescience/greenhouse-effect

**Boston University**

"*A simple greenhouse effect model*

  A. *Glass represents the 'normal' greenhouse effect on earth and is at top of atmosphere*
  B. *Solar shortwave radiation S largely makes it to surface*
  C. *For energy balance, top of glass must send S back out*
  D. *Greenhouse gases don't have a preferred direction; they send S units in both directions – up and down*
  E. *Thus, the surface of the earth receives 2S due to the greenhouse effect – instead of 1S if there were no atmosphere!*
  F. *Thermal radiation emitted from earth = 2S* "

http://people.bu.edu/nathan/ge510_06_6.pdf

# References

[1] "Joseph Fourier," [Online]. Available: http://en.wikipedia.org/wiki/Joseph_Fourier.

[2] "Horace-Bénédict de Saussure," [Online]. Available: http://en.wikipedia.org/wiki/Horace-B%C3%A9n%C3%A9dict_de_Saussure.

[3] "Translation by W M Connolley of: Fourier 1827: MEMOIRE sur les temperatures du globe terrestre et des espaces planetaires," [Online]. Available: http://www.wmconnolley.org.uk/sci/fourier_1827/fourier_1827.html.

[4] G. J. V. Wylen, Thermodynamics, John Wiley & Sons, 1960.

[5] "Climate and Climate Change," [Online]. Available: http://www.atmos.washington.edu/2002Q4/211/notes_greenhouse.html.

[6] J. Kiehl and K. Trenberth, "Earth's Annual Global Mean Energy Budget," *Bulletin of the American Meteorological Society,* no. 78(2), pp. 197-208, 1997.